\newcommand{\HI}{H${\rm\scriptstyle I}$}
\newcommand{\la}{\lesssim}
\newcommand{\ga}{\gtrsim}
\newcommand{\Ho}{H$^{\rm\scriptstyle o}$}
\newcommand{\HII}{H$^+$}
\newcommand{\Ha}{H$\alpha$}
\newcommand{\Hb}{H$\beta$}

\newcommand{\aap}{AAp}
\newcommand{\apj}{ApJ} 
\newcommand{\aj}{AJ} 
\newcommand{\apjs}{ApJS}
\newcommand{\apjl}{ApJL} 
\newcommand{\mnras}{MNRAS}
\newcommand{\Msun}{M$_\odot$} 
\newcommand{\kms}{$\,\rm {km\,s^{-1}} $} 

\documentclass[natbib]{iaus}
\usepackage{graphicx,natbib}

\title[Warm gas accretion onto the Galaxy] {Warm gas accretion onto the Galaxy}

\author[J. Bland-Hawthorn]{J. Bland-Hawthorn}

\affiliation{School of Physics, University of Sydney, Australia, NSW 2006 \\ email: {\tt jbh@physics.usyd.edu.au} \\[\affilskip]
}

\pubyear{2008}
\volume{254} 
\jname{The Galaxy Disk in Cosmological Context}
\editors{J. Andersen, J. Bland-Hawthorn \& B. Nordstrom, eds.}
\begin{document}

\maketitle

\begin{abstract}
We present evidence that the accretion of warm gas onto the Galaxy today is
at least as important as cold gas accretion. For more than a decade, the 
source of the bright \Ha\ emission (up to 750
mR\footnote{1~Rayleigh (R) =  $10^6/4\pi$ photons cm$^{-2}$ s$^{-1}$ sr$^{-1}$,
equivalent to $5.7\times 10^{-18}$ erg cm$^{-2}$ s$^{-1}$ arcsec$^{-2}$ at \Ha.})
along the Magellanic Stream has remained a mystery. 
We present a hydrodynamical 
model that explains the known properties of the \Ha\  emission and provides 
new insights on the lifetime of the Stream clouds. The upstream clouds are 
gradually disrupted due to their interaction with the hot halo gas. 
The clouds that follow plough into gas ablated from the upstream clouds, 
leading to shock ionisation at the leading edges of the downstream clouds. Since 
the following clouds also experience ablation, and weaker \Ha\ (100$-$200 mR)
is quite extensive, a disruptive cascade must be operating along much of the Stream. 
In order to light up much of the Stream as observed, it must have a small angle of attack 
($\approx 20^\circ$) to the halo,
and this may already find support in new \HI\  observations. Another prediction is that the
Balmer ratio (\Ha/\Hb) will be substantially enhanced due to the slow shock; this will soon
be tested by upcoming WHAM observations in Chile. We find that the clouds are evolving on timescales of 100$-$200~Myr, such that the
Stream must be replenished by the Magellanic Clouds at a fairly constant rate 
($\gtrsim$ 0.1 M$_\odot$ yr$^{-1}$). The ablated material falls onto the Galaxy as a warm drizzle; diffuse ionized gas
at 10$^4$K is an important constituent of galactic accretion. The observed \Ha\ emission 
provides a new constraint on the rate of disruption of the Stream and, consequently, the infall 
rate of metal-poor gas onto the Galaxy. We consider the stability of \HI\ clouds falling towards
the Galactic disk and show that most of these must break down into smaller fragments that become
partially ionized.  The Galactic halo is expected to have huge numbers of smaller neutral and
ionized fragments. When the ionized component of the infalling gas is
accounted for, the rate of gas accretion is $\sim$0.4 M$_\odot$ yr$^{-1}$, roughly twice the rate
deduced from \HI\ observations alone. 

\end{abstract}


\section{Introduction}
It is now well established that the observed baryons over the electromagnetic spectrum account for only a fraction of the expected baryon content in Lambda Cold Dark Matter cosmology. This is true on scales of galaxies and, in particular, within the Galaxy where easily observable phases have been studied in great detail over many years. The expected baryon fraction ($\Omega_B/\Omega_{DM}\approx 0.17$) of the dark halo mass ($1.4\times 10^{12}$ M$_\odot$; Smith \etal\ 2007) leads to an expected baryon mass of $2.4\times 10^{11}$ M$_\odot$ but a detailed inventory reveals only a quarter of this mass (Flynn \etal\  2006\footnote{A decade ago, it was claimed that MACHOs may
be important in the halo but these can only make up a negligible fraction by mass (Tisserand \etal\ 2007).}). Moreover, the build-up of stars in the Galaxy requires an accretion rate of $1-3$ M$_\odot$ yr$^{-1}$ (Williams \& McKee 1997; Binney \etal\ 2000), substantially larger than what can be accounted for from direct observation. We can extend the same 
argument to M31 where the total baryon mass is $\la10^{11}$ M$_\odot$ 
(Tamm \etal\ 2007). For the Galaxy, the predicted baryon mass may be a lower bound if the upward correction in the LMC-SMC orbit motion reflects a larger halo mass (Kallivayalil \etal\ 2006; Piatek \etal\ 2008; cf. Wilkinson \& Evans 1999). Taken together, these statements suggest that most of the baryons on scales of galaxies have yet to be observed.

So how do galaxies accrete their gas? Is the infalling gas confined by dark matter?
Does the gas arrive cold, warm or hot? Does the gas rain out of the halo onto the disk or is it
forced out by the strong disk-halo interaction? These issues have never been
resolved, either through observation or through numerical simulation. \HI\
observations of the nearby universe suggest that galaxy mergers and collisions
are an important aspect of this process (Hibbard \& van Gorkom 1996), but tidal interactions do not 
guarantee that the gas settles to one or other galaxy. The most spectacular interaction
phenomenon is the Magellanic \HI\ Stream that trails from the LMC-SMC system 
(10:1 mass ratio) in orbit about the Galaxy. Since its discovery in the 1970s, there have been repeated 
attempts to explain the Stream 
in terms of tidal and/or viscous forces (q.v. Mastropietro \etal\ 2005; Connors \etal\ 2005). 
Indeed, the Stream has become a benchmark against which to judge 
the credibility of N-body$+$gas codes in explaining gas processes in galaxies. A fully 
consistent model of the Stream continues to elude even the most sophisticated codes.

Here, we demonstrate that \Ha\ detections along the Stream (Fig. 1) are providing new insights 
on the present state and evolution of the \HI\ gas. At a 
distance of $D\approx 55$~kpc, the expected \Ha\ signal excited by the cosmic and Galactic 
UV backgrounds are about 3~mR and 25~mR respectively (Bland-Hawthorn \& Maloney 1999, 
2002), significantly lower than the mean signal of 100$-$200 mR, and much 
lower than the few bright detections in the range $400-750$ mR (Weiner, Vogel \& Williams 
2002). This signal cannot have a stellar origin since repeated attempts to detect stars along the 
Stream have failed.

Some of the Stream clouds exhibit compression fronts and head-tail morphologies 
(Br\"uns \etal\ 2005) and this is suggestive of confinement by a tenuous external medium. 
But the cloud:halo density ratio ($\eta = \rho_c/\rho_h$) necessary for confinement can be orders 
of magnitude {\it larger} than that required to achieve shock-induced \Ha\ emission 
(e.g. Quilis \& Moore 2001). Indeed,  the best estimates of the halo density at the distance of the 
Stream ($\rho_h \sim 10^{-4}$ cm$^{-3}$;  Bregman 2007) are far too tenuous to induce strong 
\Ha\ emission at a cloud face. It is therefore surprising to discover that the brightest \Ha\ 
detections lie at the leading edges of \HI\ clouds (Weiner \etal\  2002) and thus appear to 
indicate that shock processes are somehow involved.
 
We summarize a model, first presented in Bland-Hawthorn \etal\ (2007), that goes a long way towards 
explaining the \Ha\ mystery. 
The basic premise is that a tenuous external medium not only confines clouds, but
also disrupts them with the passage of time. The growth time for Kelvin-Helmholtz (KH) instabilities 
is given by $\tau_{\rm KH} \approx \lambda \eta^{0.5}/v_h$ where $\lambda$ is the wavelength of the 
growing mode, and $v_h$ is the apparent speed of the halo medium ($v_h\approx 350$ km s$^{-1}$; 
see \S 2).  At the distance of the Stream, the expected timescale for KH instabilities is less than for 
Rayleigh-Taylor (RT) instabilities (see \S 3). For cloud sizes of 
order a few kiloparsecs and $\xi \approx 10^4$, the KH timescale can be much less than an 
orbital time ($\tau_{\rm MS} \approx 2\pi D/v_h \approx 1$ Gyr).
Once an upstream cloud becomes disrupted, the fragments are slowed with respect to the 
LMC-SMC orbital speed and are subsequently ploughed into by the following
clouds. 

In \S 2, the new hydrodynamical models are described and the results are presented;  
we discuss the implications of our model and suggest avenues for future research.
In \S 3, we discuss the stability of \HI\ clouds (high velocity clouds) moving through the corona 
toward the Galactic disk and briefly consider the Smith Cloud, arguably the HVC with the best
observed kinematic and photometric parameters.

\section{A new hydrodynamical model}

There have been many attempts to understand how gas clouds interact with an ambient medium 
(Murray, White \& Blondin 1993; Klein, McKee \& Colella 1994). In order to capture the evolution of a 
system involving instabilities with large density gradients correctly, grid based methods 
(Liska \& Wendroff 1999; Agertz \etal\ 2007) are favoured over other schemes (e.g. Smooth Particle Hydrodynamics). 
We have therefore investigated the dynamics of the Magellanic Stream with two 
independent hydrodynamics codes, {\it Fyris} (Sutherland 2008) and {\it Ramses} (Teyssier 2002), that 
solve the equations of gas dynamics with adaptive mesh refinement. The results shown here are 
from the {\it Fyris} code because it includes non-equilibrium ionization, but we get comparable 
gas evolution from either code\footnote{Further details on the codes and comparative simulations are
provided at http://www.aao.gov.au/astro/MS.}.

The brightest emission is found along the leading edges of clouds MS~II, III and 
IV with values as high as 750 mR  for MS II. The \Ha\ line 
emission is clearly resolved at $20-30$ km s$^{-1}$ FWHM, and shares the same radial velocity 
as the \HI\ emission within the measurement errors (Weiner \etal\ 2002; Madsen \etal\ 2002). This provides an important constraint on the
physical processes involved in exciting the Balmer emission.

In order to explain the \Ha\ detections along the Stream, we concentrate our efforts on the 
disruption of the clouds labelled MS I$-$IV (Br\"uns \etal\ 2005). The Stream is trailing the 
LMC-SMC system in a counter-clockwise, near-polar orbit as viewed from the Sun. 
The gas appears to extend from the LMC dislodged through tidal disruption
although some contribution from drag must also be operating (Moore \& Davis 1994). Recently, 
the Hubble Space Telescope has determined an orbital velocity of 378$\pm$18 km s$^{-1}$ 
for the LMC. While this is higher than earlier claims, the result has been confirmed by 
independent researchers (Piatek \etal\ 2008). Besla \etal\ (2007) conclude 
that the origin of the Stream may no longer be adequately explained with existing numerical 
models. The Stream velocity along its orbit must be comparable to the 
motion of the LMC; we adopt a value of $v_{\rm MS} \approx 350$ km s$^{-1}$.


\begin{figure}[htbp]
\begin{center}
\includegraphics[width=4in]{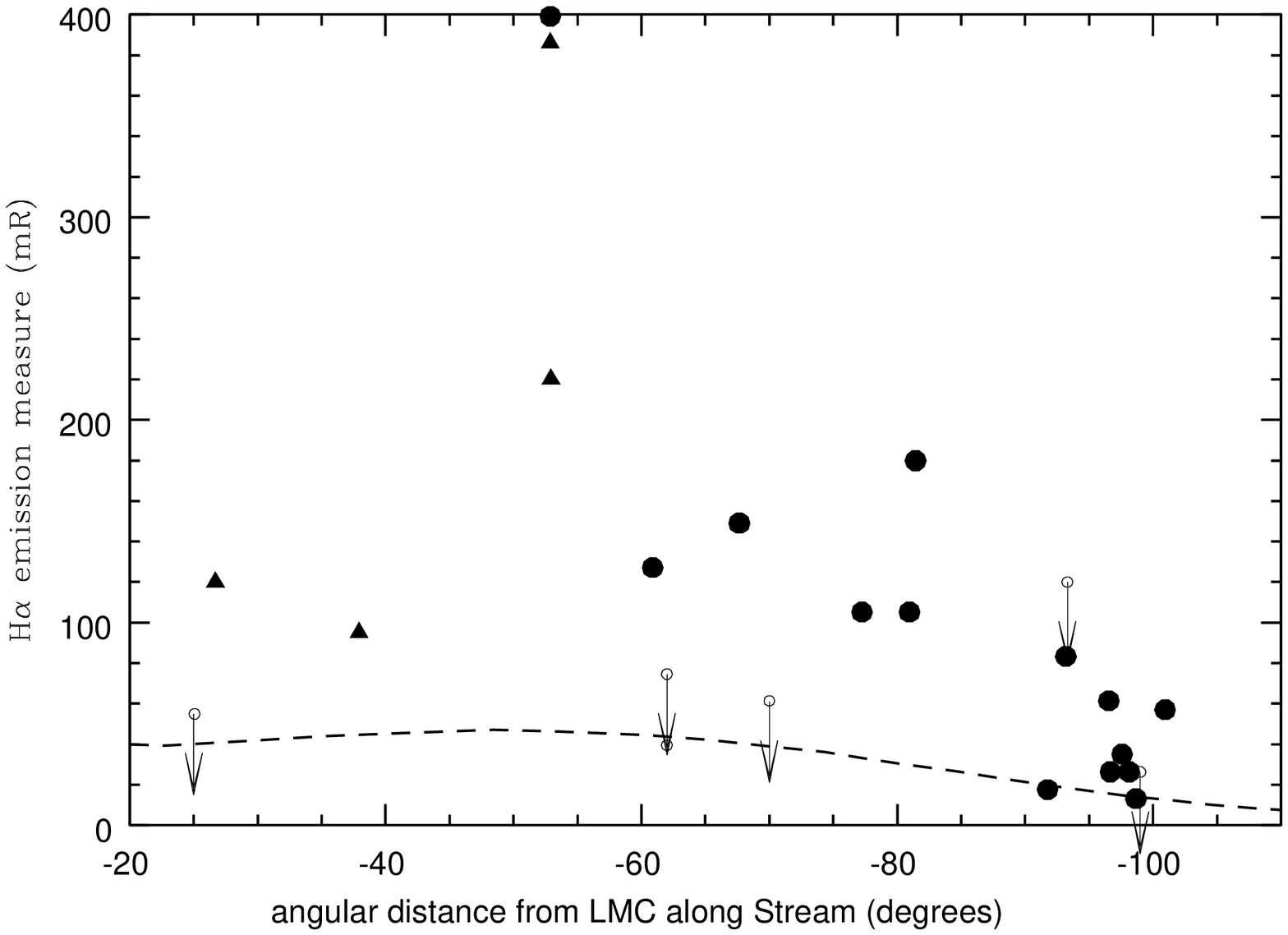}
\caption{H$\alpha$ measurements and upper limits along the Stream. The filled circles
are from the WHAM survey by Madsen \etal\ (2002); the filled triangles are from the
TAURUS survey by Putman \etal\ (2003). The dashed line model is the H$\alpha$ emission
measure induced by the ionizing intensity of the Galactic disk (Bland-Hawthorn \& Maloney
1999; 2002); this fails to match the Stream's H$\alpha$ surface brightness by at least a factor
of 3.}
\label{default}
\end{center}
\end{figure}

\begin{figure}[htbp]
\begin{center}
\includegraphics[width=4in]{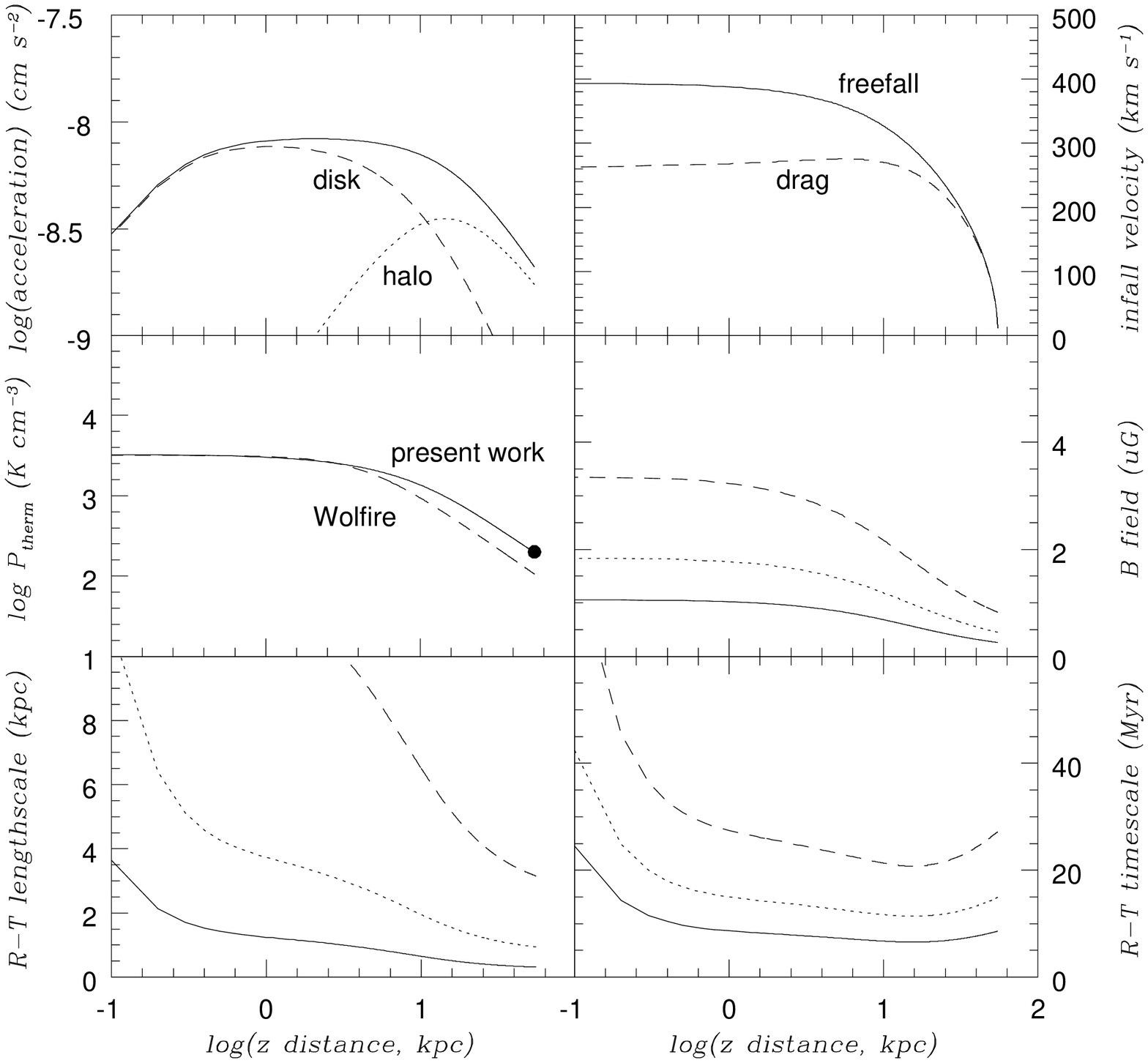}
\caption{
(a) gravitational acceleration due to the disk$+$halo $-$ all plots are shown as a function of vertical height above the disk at the Solar Circle;
(b) infall velocity for a point mass starting from the Magellanic Stream with halo drag ($C_D=1$ discussed in \S3; dashed curve) and without (solid);
(c) coronal gas pressure for our model (solid line) compared to the Wolfire model (dashed) -- the dot indicates the halo pressure used in our hydrodynamical model;
(d) magnetic field strength for $\beta=1$ (dashed), $\beta=0.3$ (dotted), $\beta=0.1$ (solid);
(e) minimum lengthscale for RT instability (discussed in \S 3);
(e) timescale for RT instability (discussed in \S 3).
}
\label{default}
\end{center}
\end{figure}
\subsection{Model parameters}

Here we employ a 3D Cartesian grid with dimensions $18\times 9 \times9$~kpc [$(x,y,z) = 
(432, 216, 216)$ cells] to model a section of the Stream where $x$ is directed along the Stream arc and
the $z$ axis points towards the observer. The grid is initially filled with two gas components. 
The first is a hot thin medium representing the halo corona.  

Embedded in the hot halo is (initially) cold \HI\ material with a total \HI\ mass of 
$3\times 10^7$ M$_\odot$. The cold gas has a fractal distribution and is initially confined to a cylinder with 
a diameter of 4 kpc and length 18 kpc (Fig. 5); the mean volume and column densities are 
0.02 cm$^{-3}$ and $2\times 10^{19}$ cm$^{-2}$ respectively. The 3D spatial power spectrum 
($P(k)\propto k^{-5/3}$) describes a Kolmogorov turbulent medium with a minimum wavenumber $k$ 
corresponding to a spatial scale of 2.25 kpc, comparable to the size of observed clouds along the Stream.

We consider the hot corona to be an isothermal gas in hydrostatic equilibrium with
the gravitational potential, $\phi(R,z)$, where $R$ is the Galactocentric radius and
$z$ is the vertical scale height. We adopt a total potential of the form $\phi=\phi_d
+\phi_h$ for the disk and halo respectively; for our calculations at the Solar Circle, 
we ignore the Galactic bulge. The galaxy potential is defined by
\begin{eqnarray}
\phi_d (R,z) &=& -c_d v_{circ}^2 / (R^2+(a_d+\sqrt{z^2+b_d^2} )^2)^{0.5}\\
\phi_h (R,z) &=& c_h v_{circ}^2 \ln ((\psi-1)/(\psi+1))
\end{eqnarray}

\smallskip\noindent
and $\psi = (1+(a_h^2+R^2+z^2)/r_h^2)^{0.5}$. The scaling constants are
$(a_d,b_d,c_d) = (6.5,0.26,8.9)$ kpc and $(a_h,r_h) = (12,210)$ kpc with $c_h=0.33$
(e.g. \citealt{miya75,wolf95}). The circular velocity $v_{circ} \approx 220$ km s$^{-1}$
is now well established through wide-field stellar surveys (\citealt{smith07}).

We determine the vertical acceleration at the Solar Circle 
using $g = -\partial\phi(R_o,z)/\partial z$
with $R_o=8$ kpc. The hydrostatic halo pressure follows from
\begin{equation}
{{\partial \phi}\over{\partial z}} = -{{1}\over{\rho_h}} {{\partial P}\over{\partial z}}
\end{equation}
After \citet{ferrara94}, we adopt a solution of the form $P_h(z) = P_o \exp((\phi(R_o,z)-\phi(R_o,0))/\sigma_h^2)$
where $\sigma_h$ is the isothermal sound speed of the hot corona. To arrive at $P_o$,
we adopt a coronal halo density of $n_{e,h} = 10^{-4}$ cm$^{-3}$ at the Stream distance (55 kpc) in 
order to explain the Magellanic Stream H$\alpha$ emission \citep{jbh07}, although
this is uncertain to a factor of a few. We choose $T_h = 2\times 10^6$ K
to ensure that OVI is not seen in the diffuse corona consistent with observation
\citep{sembach03}; this is consistent with a rigorously isothermal halo for the Galaxy.
Our solution to equation (2.3) is shown in Fig. 2(c) and it compares 
favorably with the pressure profile derive by others (e.g. \citealt{wolf95,stern02}).

A key parameter of the models is the ratio of the cloud to halo pressure, $\xi = P_c/P_h$. 
If the cloud is to survive the impact of the hot halo, then $\xi \gtrsim 1$. A shocked cloud is 
destroyed in about the time it takes for the internal cloud shock to cross the cloud, during which time 
the cool material mixes and ablates into the gas streaming past. Only massive clouds with dense cores
can survive the powerful shocks. An approximate lifetime\footnote{Here we correct a typo in equation (1) of Bland-Hawthorn \etal\ (2007).} for a spherical cloud of diameter $d_c$ is
\begin{equation}
\tau_{\rm c} = 60  (d_c/2 {\rm\; kpc}) ( v_{\rm h}/350 {\rm\; km s}^{-1})^{-1} (\eta/100)^{0.5} \,  {\rm Myr}.
\end{equation}
For $\eta$ in the range of 100$-$1000, this corresponds to 60$-$180 Myr for individual clouds. 
With a view to explaining the \Ha\ observations, we focus our simulations on the lower end of this range.

For low $\eta$, the density of the hot medium is $n_h=2\times 10^{-4}$ cm$^{-3}$.
The simulations are undertaken in the frame of the cold \HI\ clouds, so the halo gas is given an initial 
transverse velocity of 350\kms. The observations reveal that the mean \Ha\ emission has a slow trend
along the Stream which requires the Stream to move through the halo at a small angle of
attack (20$^o$) in the plane of the sky (see Fig. 5). Independent evidence for this appears to come
from a wake of low column clouds along the Stream (Westmeier \& Koribalski 2008).
Thus, the velocity of the hot gas as seen by the Stream is
$(v_x,v_y)$ $=$ $(-330,-141)$\kms. The adiabatic sound speed of the halo gas is 
200\kms, such that the drift velocity is mildly supersonic (transsonic), with a Mach number of 1.75.

A unique feature of the {\it Fyris} simulations is that they include non-equilibrium cooling through 
time-dependent ionisation calculations (cf. Rosen \& Smith 2004). When shocks occur within the 
inviscid fluid, the jump shock conditions are solved across the discontinuity. This allows us to 
calculate the Balmer
emission produced in shocks and additionally from turbulent mixing along the Stream (e.g. Slavin \etal\ 1993). 
We adopt a 
conservative value for the gas metallicity of [Fe/H]=-1.0 (cf. Gibson \etal\ 2000); a higher value 
accentuates the cooling and results in denser gas, and therefore stronger \Ha\ emission along the
Stream.



\begin{figure}[htbp]
\begin{center}
\caption{The dependence of the evolving fractions of \Ho\ and \HII\ on column density as the shock cascade
progresses. The timesteps are 70 (red), 120 (magenta), 170 (blue), 220 (green) and 270 Myr (black).
The lowest column \HI\ becomes progressively more compressed with time but the highest column \HI\ is 
shredded in the cascade process; the fraction of ionized gas increases with time. The pile-up of electrons
at low column densities arises from the x-ray halo.}
\label{default}
\end{center}
\end{figure}

\subsection{Results}

The results of the simulations are shown in Figs. $3-5$; we provide animations of
the disrupting stream at http://www.aao.gov.au/astro/MS. In our
model, the fractal Stream experiences a ``hot wind" moving in the
opposite direction. The sides of the Stream clouds are
subject to gas ablation via KH instabilities due to the reduced
pressure (Bernouilli's theorem). The ablated gas is slowed
dramatically by the hot wind and is transported behind the cloud.
As higher order modes grow, the fundamental mode associated
with the cloud size will eventually fragment it. The ablated gas
now plays the role of a ``cool wind" that is swept up by the pursuing
clouds leading to shock ionization and ablation of the downstream
clouds. The newly ablated material continues the trend along the
length of the Stream. The pursuing gas cloud transfers momentum
to the ablated upstream gas and accelerates it;
this results in Rayleigh-Taylor (RT) instabilities, especially at the stagnation point in the 
front of the cloud. We rapidly approach a nonlinear regime where the KH and RT
instabilities become strongly entangled, and the internal motions become highly turbulent.
The simulations track the progression of the shock fronts as they propagate into the cloudlets.

In Fig. 3, we show the predicted conversion of neutral to ionized hydrogen due largely to cascading
shocks along the Stream. The drift of the peak to higher columns is due to the shocks eroding away the outer
layers, thereby progressing into increasingly dense cloud cores. The ablated gas drives a shock 
into the \HI\ material with a shock speed of $v_s$ measured in the cloud 
frame. At the shock interface, once ram-pressure equilibrium is reached, we find
$v_s \approx v_{\rm h} \eta^{-0.5}$. In order to produce significant \Ha\ emission, $v_s \gtrsim 35$\kms\ such
that $\eta \lesssim 100$. In Fig.~4, we see the steady rise in \Ha\ emission along the Stream, reaching $100-200$
mR after 120~Myr, and the most extreme observed values after 170~Myr. The power-law
decline to bright emission measures is a direct consequence of the shock cascade. The shock-induced 
ionization rate is $1.5\times 10^{47}$ phot s$^{-1}$ kpc$^{-1}$. The predicted luminosity-weighted line 
widths of 20\kms\ FWHM (Fig. 4, inset) are consistent with the \Ha\ kinematics.
In Fig. 5, the \Ha\ emission is superimposed onto the projected \HI\ emission: much of it lies at 
the leading edges of clouds, although there
are occasional cloudlets where ionized gas dominates over the neutral column. Some of the brightest 
emission peaks appear to be due to limb brightening, while others arise from chance alignments.

The simulations track the degree of turbulent 
mixing between the hot and cool media brought on by KH instabilities (e.g. Kahn 1980). The turbulent 
layer grows as the flow develops, mixing up hot and cool gas at a characteristic temperature of 
about 10$^4$K. In certain situations, a sizeable \Ha\ luminosity can be generated (e.g. Canto \& 
Raga 1991) and the expected line widths are comparable to those observed in the Stream. 
Indeed, the simulations reveal that the fractal clouds develop a warm ionized skin along the entire
length of the Stream. But the characteristic \Ha\ emission (denoted by the shifting peak in Fig. 4)
is comparable to the fluorescence excited by the 
Galactic UV field (Bland-Hawthorn \& Maloney 2002). We note with interest that narrow Balmer lines can
arise from pre-cursor shocks (e.g. Heng \& McCray 2007), but these require conditions that are unlikely 
to be operating along the Stream.


\begin{figure}[htbp]
\begin{center}
\caption{The evolving distribution of projected \Ha\ emission as the shock cascade progresses. 
The timesteps are explained in Fig. 3.
The extreme emission measures increase with time and reach the observed mean values after 120~Myr; this trend in brightness arises because denser material is ablated as the cascade
evolves.
The mean and peak emission measures along the Stream are indicated, along with the approximate 
contributions from the cosmic and Galactic UV backgrounds.
{\bf Inset:} The evolving \Ha\ line width as the shock cascade progresses; the velocity scale is with respect
to the reference frame of the initial \HI\ gas. The solid lines are flux-weighted line profiles; the dashed lines are 
volume-weighted profiles that reveal more extreme kinematics at the lowest densities.}
\label{default}
\end{center}
\end{figure}

\begin{figure}[htbp]
\begin{center}
\caption{The initial fractal distribution of \HI\ at 20~Myr, shown in contours, before the wind action has taken hold. The upper figure is in the $x-z$ frame as seen from above; the lower figure is the projected distribution on the plane 
of the sky ($x-y$ plane). Both distributions are integrated along the third axis. The logged \HI\ contours correspond to 18.5 (dotted), 19.0, 19.5, 20.0, and 20.5 (heavy) cm$^{-2}$. The greyscale shows weak levels of \Ha\ along
the Stream where black corresponds to 300~mR.
The predicted \HI\ (contours) and \Ha\  (greyscale) distributions after 120~Myr. 
The angle of attack in the [x,y,z] coordinate frame is indicated. The \Ha\ emission is 
largely, but not exclusively, associated with dense \HI\ gas.} 
\label{default}
\end{center}
\end{figure}

\subsection{Discussion}

We have seen that the brightest \Ha\ emission along the Stream can be understood in terms of shock 
ionization and heating in a transsonic (low Mach number) flow. For the first time, the Balmer emission (and 
associated emission lines) provides diagnostic information at any position along the Stream that is 
independent of the \HI\  observations. Slow Balmer-dominated shocks of this kind (e.g. Chevalier
\& Raymond 1978) produce partially ionized media where a significant fraction of the \Ha\ emission
is due to collisional excitation. This can lead to Balmer decrements (\Ha/\Hb\ ratio) in excess of 4, i.e. 
significantly enhanced over the pure recombination ratio of about 3,
that will be fairly straightforward to verify in the brightest regions of the Stream.

The shock models predict a range of low-ionization emission
lines (e.g. OI, SII), some of which will be detectable even though suppressed by the low gas-phase
metallicity. There are likely to be EUV absorption-line diagnostics through the shock interfaces
revealing more extreme kinematics (Fig. 4, inset), but these detections (e.g. OVI) are only possible 
towards fortuitous background sources (Sembach \etal\ 2001; Bregman 2007). The predicted 
EUV/x-ray emissivity from the post-shock regions is much too low to be detected in emission.

The characteristic timescale for large changes is roughly 100$-$200 Myr,
and so the Stream needs to be replenished by the outer disk of the LMC at a fairly constant rate
(e.g. Mastropietro \etal\ 2005). The timescale can be extended with larger $\eta$ values (equation (2.4)), 
but at the expense of substantially diminished \Ha\ surface brightness. In this respect, we consider $\eta$ 
to be fairly well bounded by observation and theory.

What happens to the gas shedded from the dense clouds? Much of the diffuse gas will
become mixed with the hot halo gas suggesting a warm accretion towards the inner Galactic halo. 
If most of the Stream gas enters the Galaxy via this process, the derived gas accretion rate is
$\sim 0.4$\Msun\ yr$^{-1}$. The higher value compared to \HI\ (e.g. Peek \etal\ 2008) is due to the 
gas already shredded, not seen by radio telescopes now. In our model, the HVCs
observed today are unlikely to have been dislodged from the Stream by the process
described here. These may have come from an earlier stage of the LMC-SMC interaction with the 
outer disk of the Galaxy.

The ``shock cascade'' interpretation for the Stream clears up a nagging uncertainty about the 
\Ha\ distance scale for high-velocity clouds.  Bland-Hawthorn \etal\ (1998) first showed that
distance limits to HVCs can be determined from their observed \Ha\ strength due to 
ionization by the Galactic radiation field, now confirmed by clouds with reliable distance 
brackets from the stellar absorption line technique (Putman \etal\ 2003; Lockman \etal\ 2008; 
Wakker \etal\ 2007).  HVCs
have smaller kinetic energies compared to the Stream clouds, and
their interactions with the halo gas are not expected to produce significant shock-induced 
or mixing layer \Ha\ emission, thereby supporting the use of \Ha\ as a crude distance indicator.

Here, we have not attempted to reproduce the \HI\ observations of the Stream in detail. 
This is left to a subsequent paper where we explore a larger parameter space and include a more 
detailed comparison with the \HI\ and \Ha\ power spectrum, inter alia. We introduce additional physics, 
in particular, the rotation of the hot halo, a range of Stream orbits through the halo gas, and so on.

If we are to arrive at a satisfactory understanding of the Stream interaction with the halo, future 
deep \Ha\ surveys will be essential. It is plausible that current \Ha\ observations are still missing
a substantial amount of gas, in contrast to the deepest \HI\ observations. We can compare
the particle column density inferred from \HI\ and \Ha\ imaging surveys. The limiting
\HI\ column density is about  $N_H \approx \langle n_H \rangle L \approx 10^{18}$ cm$^{-2}$
where $\langle n_H \rangle$ is the mean atomic hydrogen density, 
and $L$ is the depth through the slab. By comparison, the \Ha\ 
surface brightness can be expressed as an equivalent emission measure, 
$E_m \approx \langle n_e^2 \rangle L \approx \langle n_e\rangle N_e$. Here
$n_e$ and $N_e$ are the local and column electron density. 
The limiting value of $E_m$ in \Ha\ imaging is about 100 mR,
and therefore $N_e \approx 10^{18}/\langle n_e \rangle$ cm$^{-2}$.
Whether the ionized and neutral gas are mixed or distinct, we can
hide a lot more ionized gas below the imaging threshold for a fixed $L$, particularly
if the gas is at low density ($\langle n_e \rangle \ll 0.1$ cm$^{-3}$).
A small or variable volume filling factor can complicate this picture but, 
in general, the ionized gas still wins out because of ionization of low 
density \HI\ by the cosmic UV background (Maloney 1993).
In summary, even within the constraints of the cosmic microwave background
(see Maloney \& Bland-Hawthorn 1999),
a substantial fraction of the gas can be missed if it occupies
a large volume in the form of a low density plasma (e.g. Rasmussen \etal\ 2003).

\section{Direct infall of \HI\ onto the disk}

The conditions operating along the Magellanic Stream are unlikely to be representative 
of all \HI\ clouds that move through the Galactic corona. A related process is the
infall of individual \HI\ clouds towards the Galactic disk. The Galactic halo is home to many
HVCs of unknown origin (Wakker 2001; Lockman \etal\ 2008).
The survival and stability of these clouds is a problem that has long been recognized
(e.g. Benjamin \& Danly 1997) which we now discuss.

It is likely that many or all halo clouds have experienced some deceleration during
their transit through the lower halo. Using equation (2.3), we determine a freefall velocity for
a cloud starting at the distance of the Stream (Fig. 1(b)) that is more than twice
what is inferred for clouds at the Solar Circle \citep{wakker01}, although some HVCs
clearly have high space motions (e.g. Lockman \etal\ 2008). To explain this 
observation, \citet{benjamin97} investigated a drag equation for a cloud moving through 
a stationary medium,
\begin{equation}
\mu_c \dot{v_c} = \frac{1}{2} C_D \rho_h(z) v_c^2 - \mu_c g(z)
\end{equation}
where $\mu_c$ is the surface density of the cloud. Equation (3.1) only holds as long 
as the cloud stays together. The drag coefficient $C_D$ is a measure of the efficiency 
of momentum transfer to the cloud. For the high Reynolds numbers typical of astrophysical
media, incompressible objects have $C_D\approx 0.4$ (e.g. a rough sphere) 
which indicates that the turbulent wake behind the plunging object efficiently transfers 
momentum to the braking medium. The leading face of a compressible cloud may become 
flattened, such that the approaching medium is brought to rest in the reference frame of
the cloud; in this instance, 
$C_D \ga 1$ may be more appropriate (we adopt $C_D = 1$ here).

A solution
specific to our model is shown in Fig. 2(b) where the freefall velocity is now slowed by
about 35\%. In practice, the cloud's projected motion can be considerably less than its 
3D space velocity (e.g. \citealt{lockman08}).
In all likelihood, infalling HVCs have experienced
significant deceleration through ram pressure exerted by the corona. But
even before the cloud reaches terminal velocity,
the cloud is expected to break up (Murray \& Lin 2004).


So how do clouds resist the destructive forces of RT and shock instabilities?
In \S 3.1, we investigate the stabilizing influence of magnetic fields when
a cloud passes through a magnetized medium. The halo 
magnetic field is poorly constrained at the 
present time (e.g. \citealt{sun08}). We describe the uniform magnetic field in terms of the 
pressure of the halo medium, or 
\begin{equation}
{B^2 \over 8 \pi}  = \beta P_h 
\label{eqn:Bequi}
\end{equation}
such that $B \approx 1$ $\mu$G at the distance of the Stream (55 kpc) if the field is in full
equipartition with the corona (see Fig. 1(d)). But there is evidence
that the field is weaker than implied here ($\beta \approx 0.3$; \citealt{sun08}), at
least within 5 kpc of the Galactic plane
For the warm, denser low-latitude gas (Reynolds layer), we adopt the new 
parameter fits of \citet{gaensler08} 
from a re-analysis of pulsar data. The lower $\beta$ value
finds support from recent magnetohydrodynamic simulations of the Reynolds layer \citep{hill08}.

\subsection{Stability limits and growth timescales}


We consider the surface of a high velocity cloud as a boundary
between two fluids.  In practice, the Galactic ionizing radiation field 
imparts a multiphase structure to the cloud.
At all galactic latitudes within the Stream distance, HVCs with column
densities of order $10^{20}$ cm$^{-2}$ or higher have partially
ionized skins to a column depth of roughly 10$^{19}$ cm$^{-2}$ for sub-solar
gas due to the Galactic ionizing field (see \citealt{jbh99,wolf03}). Between the warm 
ionized skin and the cool inner regions is a warm neutral medium of twice the 
skin thickness; both outer layers have a mean particle temperature of $\la 10^4$K.

The cloud is denser than the halo gas.  Because of the gravitational field,
RT instabilities can grow on the boundary.
Furthermore, KH instabilities may also develop due to
the relative motion of the cloud with respect to an external medium.
Recent work has shown that 
buoyant bubbles in galaxy clusters are stabilized
against RT and KH instabilities
by viscosity and surface tension due to magnetic fields in the boundary
\citep{deyoung03,kaiser05,jones05}.  
Here we examine whether HVC boundaries are similarly stabilized 
against disruption in the Galactic halo.

When there is no surface tension, no viscosity and no
relative motion between the two media, the growth rate
of the RT instability for a perturbation
with wavenumber $k$ is $\omega = \sqrt{gk}$,
where $g$ is the gravitational acceleration at the fluid boundary.
The wavenumber is related to a perturbation length scale, $\ell = 2\pi/k$.
The instability requires a few e-folding timescales to fully develop; the
timescale is given by
\begin{equation}
t_{grow} = \omega^{-1} =  \sqrt{\ell \over 2 \pi g}.
\label{eqn:tgrow}
\end{equation}

In the presence of a magnetic field, the transverse component  ($B_{tr}$) provides
some surface tension which can help to suppress RT instabilities below a
lengthscale of
\begin{equation}
\ell_{min} = {B_{tr}^2 \over 2 \rho_c g}
\label{eqn:lmin}
\end{equation}
\citep{chandra61}.
Here $\rho_c$ is the mass density of the denser medium,
i.e. the cloud, and $B_{tr}$ is the average value of the 
transverse magnetic field at the boundary.

In order to illustrate when RT instabilities become important,
we assume a flat rotation curve for the Galaxy (e.g. \citealt{binney97})
\begin{equation}
g  \approx {v_{circ}^2 \over R} = 1.6 \times 10^{-8}
  \left({v_{circ} \over 220~{\rm km~s^{-1}} }\right)^{2}
  \left({R \over R_o} \right)^{-1}~{\rm cm~s^{-2}} .
\label{eqn:g}
\end{equation}
This is only a rough approximation to the form expected from
equations (2.1) and (2.2). We stress that the actual behaviour discussed
below, and shown in
Figs. 2(e) and (f), solves for the gravitational potential correctly.

Shortly after the 
discovery of HVCs, it was thought that they may be
self-gravitating. But this would place them at much greater distances than
the Magellanic Stream (e.g. Oort 1966) which is now known not to be the
case (e.g. Putman \etal\ 2003). Instead, we consider two cases: (i) HVCs 
in pressure equilibrium with the coronal
gas; (ii) HVCs with parameters fixed by direct observation. In (i), because the 
temperature is not strongly dependent on radius,
but the number density decreases rapidly with increasing radius, we
expect the increased pressure to compress the clouds at lower latitudes.

We estimate the impact of RT instabilities using equations (2.3) and (3.3):
for a cloud
temperature of $T_c = 10^4$K (see \S2) in pressure equilibrium with the 
hot halo, the  electron density is given by
\begin{eqnarray}
\label{eqn:nec}
n_{e,c} & \approx & n_{e,h} {T_h \over T_c}  \\
& = & 0.02 
\left({R \over 55~{\rm kpc}}\right)^{-2}
\left({T_h \over 2\times 10^6 {\rm K}}\right)
\left({T_c \over 10^4 {\rm K}}\right)^{-1}~{\rm cm^{-3}} . \nonumber
\end{eqnarray}


We use equations \ref{eqn:lmin}, \ref{eqn:nec} and  \ref{eqn:Bequi}
to estimate $\ell_{min}$ as a function of Galactocentric
radius. The minimum length scale for instability is
\begin{eqnarray}
{\ell_{min} \over R } & \sim & { 8 \pi \beta k_B T_c  \over m_p v_{circ}^2} \\
 & = & 0.004 \left({T_c \over 10^4{\rm K}}\right)
         \left({\beta \over 0.1}\right)
         \left({v_{circ} \over 220~{\rm km~s^{-1}}}\right)^{-2} \nonumber
\end{eqnarray}
and its associated growth timescale using equation (\ref{eqn:tgrow})
\begin{eqnarray}
t_{grow} & \sim & \sqrt{ 4 \beta k_B T_c  \over m_p v_{circ}^2}\;  \Omega^{-1} ~~~~~~ \\
& = &   1.1~{\rm Myr} 
      \left({T_c \over 10^4{\rm K}}\right)^{1\over2}
      \left({\beta  \over 0.1}\right)^{1 \over 2}
      \left({v_{circ} \over 220 ~{\rm km~s^{-1}}}\right)^{-2}
      \left({R \over R_o}\right) \nonumber
\end{eqnarray}
where the angular rotation rate is given by $\Omega = v_{circ}/R$.

Because we have assumed that $B^2 \propto n_{e,h}$  (equipartition)
and $n_{e,c} \propto n_{e,h}$ (pressure equilibrium),
neither the minimum scale length or its growth timescale
depend on the halo density or temperature.
They do depend on the temperature of the clouds and the
ionization state.
If the clouds are hotter than $10^4$K, then $n_{e,c}$ is overestimated
under the assumption of pressure
equilibrium.  This would
lead to larger minimum instability lengthscales
and growth timescales.
If the Galactic rotation curve drops faster than
the flat profile implied by equation (2.3),  we would have underestimated both
the minimum instability scale length and its associated
growth timescale at large radii. 

Under the assumption of pressure equilibrium, the falling clouds
become more compressed as they approach the disk which can hasten cooling.
This effect may help to stabilize against break up, particularly if a cool shell
develops (cf. Sternberg \& Soker 2008).

In the absence of gravitational instability, the flow
is stable against the KH instability if \citep{chandra61}
\begin{equation}
U^2 < {B_{tr}^2 (\rho_c + \rho_h) \over 2 \pi \rho_c \rho_h}
\end{equation}
where $U$ is  the relative velocity between the two fluids.
When $\rho_c > \rho_h$, this requirement becomes
\begin{eqnarray}
U & < &\sqrt{8 \beta k_B T_h \over m_p} \\
& = & 115  \left({\beta \over 0.1 }\right)^{1 \over 2}
   \left({T_h \over 2 \times 10^6{\rm K} }\right)^{1 \over 2}~{\rm km~s^{-1}} 
\nonumber
\end{eqnarray}
and we have described the magnetic field in terms
of the halo pressure using equation \ref{eqn:Bequi}.
This requirement is also independent of the halo density 
as we have related the magnetic field to the halo pressure,
although it is dependent on the halo temperature.
This requirement is nearly satisfied for HVCs if
the magnetic field is near equipartition. 

\subsection{The Smith Cloud}

Arguably, the high-latitude \HI\ cloud that we know most about is the
Smith Cloud. Lockman \etal\ (2008) have recently published spectacular
\HI\ data for this HVC and deduce a remarkable amount about its past
and future properties. The HVC has an estimated 
distance of $12.4\pm 1.3$ kpc, a Galactocentric radius of $R \approx 8$ kpc, a 
vertical height below the plane of -$2.9$ kpc,
a mass of at least $10^6~M_\odot$ in a volume of order 3~kpc$^3$ 
corresponding to $n_{c} \approx 0.014~{\rm cm}^{-3}$. The cloud has
a prograde orbit that is inclined 30$^\circ$ to the plane and appears to have come 
through the disk 70 Myr
ago at $R\approx 13$ kpc moving from above to below the plane.

In order to have punched through the disk, the shock crossing time for the cloud
must be longer than for the disk. It can be shown that
\begin{equation}
{{d_c}\over{z_d}} > \sqrt{{n_d}\over{n_c}}
\label{eq:punch}
\end{equation}
where $z_d$ is the vertical thickness and $n_d$ is the mean density of the \HI\ at the crossing point.
This is essentially a statement that the surface density of the cloud must be higher
than the disk. If we assume the cloud punched through the Galactic hydrogen density 
profile determined by Kalberla \& Dedes (2008), 
equation~\ref{eq:punch} indicates that the cloud was substantially
thicker than the disk when it came through and somewhat more massive than what
is observed now.  Consistent with this picture, the observed wake may result from 
ablation processes induced by the impact. For cloudlets smaller than 100 pc, thermal 
conduction due to the halo corona (McKee \& Cowie 1977) and the Galactic radiation 
field convert the ablated gas to a clumpy plasma.


The kinetic energy of the Smith Cloud observed 
today is $\sim 10^{54}$ erg -- this is enough to punch through the disk if sufficiently 
concentrated. Impulsive
shock signatures at UV to x-ray wavelengths will have largely faded away, and the
\HI\ ``hole'' at the crossing point will have been substantially stretched by
differential shear\footnote{It is sometimes claimed, this meeting notwithstanding, that 
outer disk \HI\ ``holes'' are 
evidence of dark matter minihalos passing through the disk, but it can be shown
that the gravitational impulse has negligible impact on the gaseous disk.}.

Figs. 2(e) and (f) show that a cloud of several kpc can survive RT instabilities at
these latitudes, but it is difficult to see how the Smith Cloud, like several other
large HVCs, could have come in from, say, the distance of the Magellanic Stream.
Lockman \etal\ (2008) use essentially the same Galactic potential as described
here to determine the cloud's orbit parameters. We conjecture that either the 
cloud has been dislodged from the outer disk by a passing dwarf, or the cloud
has been brought in by a confining dwarf potential. 
A cloud metallicity of [Fe/H]$\approx$-1 is appropriate in either scenario.
Interestingly, the impulse from the Galactic disk can cause the gas to become
dislodged from the confining dark halo or to oscillate within it.
The interloper must be on a prograde orbit which rules out some
infalling dwarfs (e.g. $\omega$Cen; Bekki \& Freeman 2007),
but conceivably implicates disrupting dwarfs like Canis Major or Sagittarius, 
assuming these were still losing gas in the recent past.

\subsection{Discussion}

In \S2, we presented evidence for a shock cascade along the Magellanic
Stream arising from the disruption of upstream clouds due to their 
interaction with the Galactic halo. Bland-Hawthorn \etal\ (2007) make firm
predictions that can be tested in future observations. A possible improvement 
is to consider
the entire Magellanic System, i.e. the influence of the LMC-SMC system 
that lies further upstream. Mastropietro \etal\ (2008) present evidence for
a strong interaction along the leading edge of the LMC; for their quoted
model parameters, it seems plausible that this results in a stand-off bowshock 
ahead of the galaxy. The cross wind over the face of the LMC could be confused 
for a starburst-driven wind from the LMC (cf. Lehner \& Howk 2007). 
In all likelihood, the LMC-SMC system creates a turbulent wake behind it which 
may impact the development of instabilities in the trailing stream.

The issue of cloud survival is highly complex.
In \S 3, we did not consider the role of viscosity in
quenching RT or KH instabilities.
Simulations have shown that viscosity does lead to stabilization 
\citep{pavlovski08} but we have not been able to 
estimate a lengthscale or a growth timescale appropriate
for our setting.  
\citet{kaiser05} show that when the density ratio 
between the two media is large, KH instabilities
fail to grow and the growth rate
of RT instabilities depends only on the properties of
higher density medium, in our case the cloud medium.
However their result, taken in the limit of 
one density much larger than the other, $\rho_2 \gg \rho_1$,
will not apply if $\nu_1 \rho_1 \gg \nu_2 \rho_2$.
Here the subscripts refer to the fluids on either
side of the boundary and $\nu$ is the kinematic viscosity.
Because diffusivity
coefficients are sensitive functions of temperature ($\propto T^{-2.5}$),
they could dampen fluid instabilities.  
Unfortunately the expected differences in temperatures between
HVCs and the halo gas (corona) suggest that 
$\nu_1 \rho_1 \gg \nu_2 \rho_2$ and thus we cannot
apply the limit used by \citet{kaiser05}.
A proper treatment is required to cover the Galactic halo setting.

Other studies have argued that the KH instability
leads to a turbulent mixing layer on the surface
and so is less destructive than the RT instability 
(e.g. \citealt{deyoung03}). At the present time, 
there are no relevant astrophysical codes that are capable of
handling mixing in a satisfactory manner. 
On the issue of magnetic stability, more sophisticated treatments using MHD
have been attempted, but the main conclusions appear to be 
contradictory (Konz \etal\ 2002; Gregori \etal\ 1999). We are not aware
of MHD codes that are sufficiently capable of answering this question
at the present time. 

Without excessive erudition, which is inappropriate for a conference
proceeding, it is difficult to mount a solid case for why hydro processes 
could ultimately save the day for HVCs. But the fact of the matter is that
fast-moving gas clouds do survive their passage through the Galactic halo.
These may be mostly shortlived entities on the road to destruction, 
suggesting that there is a largely hidden plasma component that we
have yet to fully comprehend. This will require more extensive observations
at difficult parts of the observational parameter space, matched by hydro
codes that can properly treat instabilities and mixing in a multi-phase gas.

\acknowledgments  JBH is supported by a Federation Fellowship through the Australian 
Research Council. I thank Alice Quillen and Ralph Sutherland for their role in the work presented
here. I acknowledge helpful discussions with Bob Benjamin, Chris Flynn, Bryan Gaensler, Greg 
Madsen and Mary Putman.

\end{document}